\documentstyle[preprint,aps,epsf]{revtex}

\begin{document}

\draft

\title{The multiple solutions of self-consistency condition in
Walecka model and the validity of the Brown-Rho scaling law}
\author{ Zhang Benwei, Hou Defu and  Li Jiarong  }
 \address{ Institute of
Particle Physics, Hua-Zhong Normal University, Wuhan 430079, China}
\maketitle
\date{today}

\begin{abstract}

We investigate the self-consistency condition (SCC) of  mean-field theory
in Walecka model and find that the solutions of the SCC
are multiple at high temperature and chemical potential. Using
 the effective Lagrangian approach, we study  medium effects on the
 $\omega$ meson mass by taking into account of vacuum effects. 
We show that the $ \omega $ meson mass decreases with both
temperature and chemical potential with  a  general tendency, while near the 
critical point several  $ \omega $ meson masses become degenerate due to the
multiple solutions of the SCC. We check the validity of Brown-Rho scaling law 
in this case. Finally, we calculate the thermodynamic potential and prove
that the  multiple solutions of the SCC result from a first-order phase transition of nuclear matter in the Walecka model at high temperature and  chemical potential.

\end{abstract}

\pacs {PACS numbers:  21.65. +f, 14.40. Cs, 11.10. Wx}
\newpage

Investigations of how the properties of hadrons are modified at
high temperature and/or density have attracted a lot of 
attention recently[1-12,16]. This studies can help us to better understand 
the properties of the chiral phase transition and the quark-gluon plasma 
which is expected to be created in relativistic heavy ion collisions. One of the most interesting issues concerns the vector meson masses in medium.
 The dependence of vector meson properties on the temperature and/or
the density of the nuclear medium is  far from being well understood. There
are still a lot of controversies:  the results obtained by
 different methods do not coincide. 
Some authors \cite{Hat,Asa} find the vector
meson masses decrease with temperature and/or density which is
interpreted as evidence of the partial restoration of chiral
symmetry\cite{Bro2,Kub}. For example, with effective chiral Lagrangians
and a suitable incorporation of the scaling property of QCD, it
has been shown \cite{Bro1} that the effective masses of $ \sigma, \rho, 
\omega$ mesons and nucleons satisfy the approximate in-medium 
scaling law
\begin{equation}
\frac{m^{*}_{\sigma} }{m_{\sigma}} \approx
\frac{m^{*}_{\rho}}{m_{\rho}}
\approx \frac{m^{*}_{\omega}}{m_{\omega}}
\approx \frac{m^{*}_{N}}{m_{N}}.
\end{equation}

\noindent Incontrast,  others [7,8] have proven that the vector meson masses
increase in hot and/or dense matter by using effective theories. 
In addition,, because 
the vacuum structure in hot and/or dense medium is changed, the vacuum
effect on the vector meson masses should be taken into account
[1,9-12]. Using this idea some authors[10-12] calculate the
vector meson masses also by using the effective langrangian approach
and find  the vector meson masses are reduced at either finite temperature 
or density.  In order to include the vacuum effect in medium, 
some calculations  use the Walecka model to get the effective nucleon mass 
$ M^{*} $ from the self-consistent condition (SCC) $ M^{*}=M-g_s\phi  $ 
by using  mean-field theory (MFT).  One obtains a  single solution  $ M^{*} $ 
in all regions of the parameter space. In this paper we perform
a careful and detailed investigation of these issues and 
 find some new results.    We first study the  SCC in MFT of the 
Walecka model  and find the solutions of the SCC are
multiple for some values of the thermodynamic parameters.
Then we investigate the $ \omega$ meson mass in
hot and dense matter within the imagine time formalism of
finite temperature field theory. We consider an  ${\omega NN}$ interaction and
include  vacuum effects . 
We show that the $ \omega $ meson mass decreases with 
both temperature and chemical potential with an overall tendency, while
 multiple  $\omega$ meson masses become degenerate at some
 values of the temperature and the chemical 
potential.  Finally we use our concrete numerical results
to check the Brown-Rho scaling law
 Eq.$(1)$ at different temperatures and chemical potentials and give a physical
interpretation for the multiple solutions in terms of the thermodynamic
potential.

We begin by using the Walecka model (often called QHD-I) [13]
to get the changes of the nucleon mass at finite temperature and density. 
The Lagrangian density in the Walecka model is given by
\begin{eqnarray*}
{\cal L} & = &\overline{\psi}[{\gamma}_{\mu}(i{\partial}^{\mu}-g_{\nu}V^{\mu})-
  (M-g_{s}\phi)]{\psi}
   + {\frac{1}{2}}({\partial}_{\mu}{\phi}{\partial}^{
  \mu}{\phi}-m_s^2 {\phi}^{2})\\
 & - &\frac{1}{4} F_{\mu\nu}F^{\mu\nu}+
  \frac{1}{2}m_v^2 V_{\mu}V^{\mu}+\delta {\cal L},
\end{eqnarray*}
\noindent which contains fields for baryons $p,n(\psi)$, the 
neutral scalar meson $\sigma(\phi)$ and the vector meson $\omega(V_{\mu})$ .
From  mean-field theory(MFT)[14], the self-consistency condition(SCC)
can be derived as
\begin{eqnarray}
  M^*=M-g_s \phi = M-\frac{g_s^2}{m_s^2}
  \frac{2}{{\pi}^2}{\int}p^2dp{\frac{M^{*}}{{\omega}^{*}_{N}}
  [{n_{p}}(T)+{ \overline{n}_{p} }(T) ]},
\end{eqnarray}
where $M^{*}$ is the nucleon effective mass and 
\begin{eqnarray*}
{n_{p}}(T)={\frac{1}{ e^{ {\beta}({\omega}^{*}_{N}-{\mu})}+1 } },\ 
{ \overline{n}_{p} }(T)=
{\frac { 1 } { e^{ {\beta}({\omega}^{*}_{N}+{\mu}) }+1} },\ 
 {\omega}^{*}_N = \sqrt{ p^2+M^{*2} } ,\  {\beta}={\frac{1}{T}}.
\end{eqnarray*}
This expression  indicates that the main contribution to the reduction of $M^{*}$ comes from
the effect of  the scalar $\sigma$ meson.
Taking the parameters [14] $ g^2_s=109.626$ and $ m_s=520$ MeV , we can  
get the nucleon effective mass $ M^{*} $ by numerically solving Eq.(2). The results are  depicted in Fig. 2 to Fig. 6. We can
see that  $ M^{*} $ decreases with  temperature T and  chemical potential
$\mu$,  and when  T and  $\mu$ are not very large there is only one solution 
of the  SCC.  However, as T and $\mu$ become
larger  there appear three different solutions for the nucleon effective mass.
In Fig. 2 we show the dependence of  $ M^{*} $ on the temperature T when the
chemical potential $\mu=0 $. There are three
solutions for $M^{*}$ when T ranges from  $185.7$ MeV to $186.6$ MeV. This
 very narrow range  was ignored in the previous literature and consequently 
all previous calculations obtained a single  solution of the  SCC for $M^{*}$ 
\footnote{Serot[14] once speculated that there might be several solutions of SCC for fixed T and $\mu$, but he did not give concrete  and affirmative results. }.
In Fig. 3 we draw the multiple solutions for $M^{*}$ in detail in the
range  and we can easily see the three different solutions
for $M^{*}$ in this range. For example, 
when $T=185.8$ MeV we  get $M^{*}=390.49$ MeV, $M^{*}=455.813$ MeV or
$M^{*}=683.377$ MeV as the solutions of the SCC. When
$T=186.4$ MeV we  get $M^{*}=352.134$ MeV, $M^{*}=534.399$ MeV, $M^{*}=648.193$ MeV. 
From Fig. 4 and Fig. 5 we  find the same
phenomenon in the dependence of $M^{*}$ on $T$ when $\mu=200$ MeV , but
the parameter range corresponding to multiple solutions enlarges to 
nearly $7$ MeV. In Fig. 6 we
show the $\mu$ dependence of $ M^{*}$ when $T=100$ MeV.  We find the three
solutions appear in a large range of nearly $200$ MeV.

Next we turn to  study the $\omega$ meson effective mass in hot and dense
matter. In Minkowski space the self-energy can be generally expressed as
\begin{equation}  
 {\Pi}^{\mu\nu}=F{P_{L}^{\mu\nu}}+G{P_{T}^{\mu\nu}}, 
\end{equation}
\noindent where $ P_{L}^{\mu\nu}$ and $P_{T}^{\mu\nu} $are stan\-dard
lon\-gitu\-dinal and trans\-verse pro\-jection tensors respectively, 
and are defined as
\begin{eqnarray}
P^{00}_T=P^{0i}_T=P^{i0}_T=0,  
P^{ij}_T={\delta}^{ij}-\frac{ {k^i}{k^j} } { {\bf k}^2 }, 
P^{\mu\nu}_L=\frac{ {k^{\mu}}{k^{\nu}} } 
{ {k^2} }-g^{\mu\nu}-P_{T}^{\mu\nu}.  
\end{eqnarray}
\noindent    The full propagator of the $\omega$ meson reads
\begin{equation}
  {\cal D}^{\mu\nu}=-\frac{P^{\mu\nu}_L}{k^2-m_{\omega}^2-F}
  -\frac{P^{\mu\nu}_T}{k^2-m_{\omega}^2-G}
  -\frac{k^{\mu}k^{\nu}}{{m_{\omega}^2}{k^2}}.
\end{equation}
By making  use of Eqs.(3) and (4), one can show

\begin{equation}
  F(k)=\frac{k^2}{{\bf k}^2} { {\Pi}^{00} }(k), \ 
  G(k)=-\frac{1}{2} [ {\Pi}^i_i+\frac{k^2_0} { {\bf k}^2  } {{\Pi}^{00}}(k)].   
\end{equation}
In order to get the effective mass of the $\omega$ meson, we  take the 
limit ${\bf k}{\rightarrow 0}$, and show that  $ F=G $ and $ k^2={k^0}^2-{\bf
k}^2={\omega}^2$. The effective mass of the $\omega$ meson 
in hot and dense medium is  obtained from the pole of the full propagator  
\begin{equation}
  {\omega}^2-m_{\omega}^2-{\rm Re}[ \lim_{{\bf k}\rightarrow 0} F(k)]=0.
\end{equation}

 The Lagrangian density for ${ \omega NN} $  interaction reads
\begin{equation}
{\cal L}_I=g_{\omega NN} \overline{\psi} {\gamma}_{\mu}
{\omega}^{\mu} {\psi},
\end{equation}
\noindent where ${\omega}^{\mu} $ and $ \psi $ are the $ \omega$ meson 
and nuclear fields re\-spec\-tively. 
Using the imagin\-ary-time formal\-ism of the finite tempe\-rature 
field theory,
one can obtain the self-energy for the  $ \omega $ meson  ( Fig.1) 
\begin{equation}
{\Pi}^{\mu\nu}_{\rho}=2{g_{\rho NN}^2}T \sum_{ n={-\infty} }^{+\infty}
\int \frac{d^3p}{ {2\pi}^3 } Tr [ { {\gamma}_{\mu} }(k) \frac{1}{{\gamma}^s 
p_s-M^{*}}{ {\gamma}_{\nu} }(-k) \frac{1}{ {\gamma}^s (p_s-k_s)-M^{*} } ],
\end{equation}
\noindent with $ p_0=(2n+1) \pi T i+\mu $, $ k_0=2 l\pi T i $ 
. By making use of  the mathematical expression
\begin{displaymath}
T{\sum}f(p_0=(2n+1) \pi T i + \mu ) 
=\frac{T}{2{\pi}i} \oint dp_0 f(p_0) \frac{1}{2} 
\beta \tanh [ \frac{1}{2} \beta (p_0-\mu) ],
\end{displaymath}
\noindent one can seperate  the vacuum contibutions from  the matter contributions [15]:
\begin{eqnarray*}
 T \sum f(p_0=(2n+1) \pi T i + \mu)   
=-\frac{1}{ 2{\pi}i } \int_{-i\infty+\mu+\epsilon}
^{+i\infty +\mu + \epsilon}dp_0 f(p_0)
\frac{1}{ e^{ \beta (p_0-\mu) }+1  }    \\
-\frac{1}{ 2{\pi}i } \int_{-i\infty+\mu-\epsilon}
^{+i\infty +\mu -\epsilon}dp_0 f(p_0)
\frac{1}{ e^{-\beta(p_0-\mu) }+1  }    
+\frac{1}{2{\pi}i}\int_{-i\infty}
^{+i\infty }dp_0 f(p_0-\mu).
\end{eqnarray*}
\noindent The first two terms correspond to matter contributions
as functions of temperature and chemical potential, while
the last term accounts for vacuum contributions.
  
The last term  is divergent. Using dimensional
regularization to get rid of the divergence, we obtain from Eq.(7)
\begin{eqnarray}
{ {\Pi}_{vac}^{\mu\nu} }(k)=\frac{g_{\omega NN}^2} { \pi^2 }
(k^{\mu} k^{\nu}-k^2 g^{\mu\nu} ){\omega^2} \int_0^1 dz
z(1-z)\ln[\frac{M^{*2}-{\omega}^2
z(1-z)}{M^2-m_{\omega}^2 z(1-z)}].
\end{eqnarray}
 Using the residue
theorem and the periodicity condition of $ k_0 $ in the finite
temperature field theory[15], we can then get the matter contribution,
 \begin{eqnarray*}
{ {\Pi}_{mat}^{00} }(k)
&=&-\frac{g_{\omega NN}^2} {\pi^2} \int dp \frac{p^2}
    { {\omega}^{*}_N }  [{n_{p}}(T)+{ \overline{n}_{p} }(T) ]  \\
& &    [ 2+\frac{4{\omega}^{*}_N k^0-4{\omega}^{*2}_N-k^2 }{4p {|\bf k|} }
    \ln \frac{ k^2-2k^0{\omega}_N^{*}+2p{|\bf k|} } 
    { k^2-2k^0{\omega}_N^{*}-2p{|\bf k|} } \nonumber    \\
& &+\frac{-4{\omega}^{*}_N k^0-4{\omega}^{*2}_N-k^2 }{4p {|\bf k|} }
   \ln \frac{ k^2+2k^0{\omega}_N^{*}+2p{|\bf k|} }  
   { k^2+2k^0{\omega}_N^{*}-2p{|\bf k|} } ]   ,  \\
( {\Pi}_i^i(k))_{mat}
&=&-2 \frac{g_{\omega NN}^2} {\pi^2} \int dp
\frac{p^2}{ {\omega}^{*}_N }[{n_{p}}(T)+{ \overline{n}_{p} }(T) ]  \\
& &  [      2-\frac{2M^{*}+k^2}{4p{|\bf k|} }( \ln \frac{
   k^2-2k^0{\omega}_N^{*}+2p{|\bf k|} }  
   { k^2-2k^0{\omega}_N^{*}-2p{|\bf k|} }  \nonumber  \\
& &+\ln \frac{ k^2+2k^0{\omega}_N^{*}+2p{|\bf k|} }  
{ k^2+2k^0{\omega}_N^{*}-2p{|\bf k|} }    )  ],
\end{eqnarray*}
where ${\omega}_N^{*}>\mu$.
From Eq.(7) we can obtain $F(k)$ and $G(k)$ which allow us to study
  the  dependence of $ m_{\omega} $  on the momentum $\bf k$. However,
 we are interested here
in the effective mass of the $\omega$ meson, thus we take the limit
${\bf k}\rightarrow 0 $  and find     

  \begin{eqnarray}
{\rm Re}[\lim_{ {\bf k}\rightarrow 0 } { F_{vac} }(k)]
&=&\frac{ g_{\rho NN}^2 }{\pi^2} {\omega}^2 \int_0^1 dz
   z(1-z)\ln[\frac{M^{*2}-{\omega}^2
   z(1-z)}{M^2-m_{\omega}^2 z(1-z)},     \\
{\rm Re}[\lim_{ {\bf k}\rightarrow 0 } { F_{mat} }(k)]
&=&-\frac{ 4g_{\rho NN}^2 }{ {\pi^2} } \int p^2 dp  
   { \frac{1}{ {\omega}^{*}_{N} ({\omega}^2 
   -4{\omega}^{*2}_{N}) } }       \nonumber  \\
&&[{\frac{1}{ e^{ {\beta}({\omega}^{*}_{N}+{\mu})}+1 } }+
  {\frac{1}{ e^{ {\beta}({\omega}^{*}_{N}-{\mu})}+1 } }]     
  (  { \frac{4}{3}  } p^2+2M^{*2} ).
\end{eqnarray}
Note that $\omega NN$ interaction will affect
$M^{*}$. However, as has been pointed out by other authors [16], 
 the effects
due to vector meson exchange upon $M^{*}$ are very small and  can be
neglected. This procedure is consistent with our treatment of 
the nucleon mass $M^{*}$ since $M^{*}$ is determined
only by the scalar and fermion fields within the Walecka model in MFT, 
and there will be no effect  on $M^{*}$ from vector mesons

With  $ M^{*} $  given by Eq.(2)  
we can solve Eq.(7) numerically and determine the  modified   meson mass 
  $m^{*}_{\omega}$ as a function of  temperature and chemical potential.
We choose a coupling of  $\frac{ g^2_{\omega NN} } { 4 \pi }=20.0 $
[17].
 In Fig. 2 to Fig.5  we 
show the temperature dependences of the $ \omega$ meson mass  at $\mu=0$
and $\mu=200$ MeV respectively and find that the $ \omega$ meson mass decreases
with temperature with an overall tendency.
 Fig.6  shows the chemical potential dependence of the $ \omega$
meson mass at $T=100$ MeV. The $ \omega$ meson mass 
also decreases with
increasing chemical potential $\mu$ with  a general tendency, which is 
consistent with the
result by other authors [18]. However, in contrast to the results of other authors   we   find 
 that for high  $T$
and/or $\mu$  there
may be several solutions for $ \omega$ meson masses.
Using our results we can
check the Brown-Rho scaling law Eq. (1).  We draw the curves of the  $T$ and
$\mu$ dependences of  $ \frac{M^{*}_N}{M_N}  $ and 
$ \frac{m^{*}_{\omega}}{ m_{\omega} }$  in figures from  Fig.2 to Fig.6
 We find that the scaling law works well  when T and $ \mu $ are not too
high and there is  a single  solutions for $ M^{*} $ and $ m_{\omega}^{*}$.
But  when the temperature and chemical potential are large
, the scaling law will be broken, since  there are several
different solutions for $ M^{*} $ and $ m_{\omega}^{*}$ . 

In order to  get deeper insight into
the multiple solutions of the SCC, we calculate  the thermodynamic
potential for  Walecka Model [19],
\begin{eqnarray*}
\Omega &=& \frac{1}{2} (m_s^2 {\phi}^2+m_v^2 {\bf V}^2-m_v^2 V_{0}^2 )V \\
       &-& T\sum_{ {\bf p},\lambda } [\ln(1+e^{-\beta ( {\omega}_N^{*}-\mu)
          } )+ \ln(1+e^{-\beta ( {\omega}_N^{*}+\mu ) } ) ]
\end{eqnarray*}
\noindent In this expression,  the sums run over all single-particle states labelled by 
momenta  ${\bf p}$ and intrinsic quantum number $\lambda$. From the 
thermodynamic potential one can obtain all of the  thermodynamic functions. 
As shown in  Fig. 7,  from the isotherms of the thermodynamic potential as a
function of $M^{*}$,  we find that when T is not too high, the thermodynamic
potential only has one extreme point, but as T increases to near $186$ MeV, there are three extreme points, which correspond with the three solutions for
$M^{*}$. Furthermore, when T continues to increase, there is again only
 one extreme point . The amplified region where 
the thermodynamic potential has three extreme points is depicted more clearly
in Fig. 8.  In Fig. 9 we show the phase diagram of nuclear matter
in the Walecka model. This diagram looks  quite similar to the Van Der Waals isotherm 
of gas-liquid transitions. Figures from Fig. 7 to Fig. 9 indicate explicitly 
 that   nuclear matter descripted by Walecka model has a first-order phase
transition at high temperature and density, which results in the multiple 
solutions for the nucleon and omega meson masses.

In summary, 
we have studied the nucleon effective mass  within the  Walecka model in the 
imagine time formalism of finite temperature field theory. Within the effective Lagrangian approach we investigate the medium effects on the $\omega$ meson 
mass  and include the vacuum effects.  We  show that
 $ m_{\omega}^{*}$ decreases with both temperature and chemical potential 
with an overall tendency. From the thermodynamic potential 
and the phase diagram we find  that Walecka model in MFT predicts
a first-order phase transition at high temperature and chemical potential,
which consequently  leads to  multiple solutions for $ M^{*} $ and
$ m_{\omega}^{*}$ which  results in the breaking of the Brown-Rho scaling law.
Of course our results  may depend  on the model and might only occur within
the mean-field approximation. It would be interesting to extend this work
to other models and to go beyond MFT. In this paper
 we have
only  discussed the $\omega$ meson effective mass in
medium,  but  our discussions can also be applied to the $\rho$ 
meson mass in hot and/or dense matter and we expect similar results will be 
obtained.  
\acknowledgments

We are grateful to M. E. Carrington for a careful reading of the
manuscript. This work was supported in part
 by the National Natural Science Foundation of China (NSFC)
 under Grant No.(19775017)

\vskip 1cm
\begin{center}
{\LARGE {\bf Figure Captions}}
\end{center}

\begin{description}
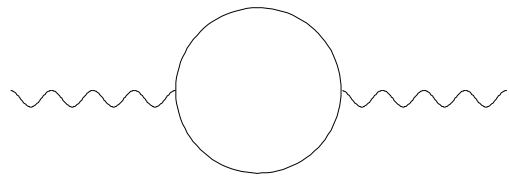
\item{{\bf Fig. 1}} One-loop diagram for $\omega$ meson self-energy in
the ${\omega NN}$ interaction.

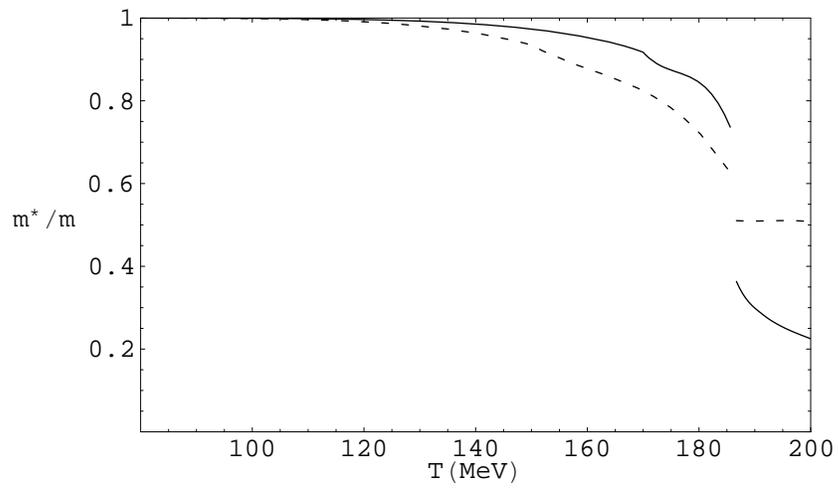
\item{{\bf Fig. 2}} The temperature dependence of
$ \frac{M^{*}_N}{M_N} $ and $\frac{ m^{*}_{\omega} }{ m_{\omega} }$ at
$\mu=0$  when there is a single solution for $M^{*}$ and $m^{*}_{\omega}$.
 The solid line represents $ \frac{M^{*}_N}{M_N} $ and the dashed line 
represents $ \frac{ m^{*}_{\omega} }{ m_{\omega} }  $. When 
T$\sim 186$ Mev each line has a cut, which means there are multiple 
solutions for   $M^{*}$ and $m^{*}_{\omega}$.

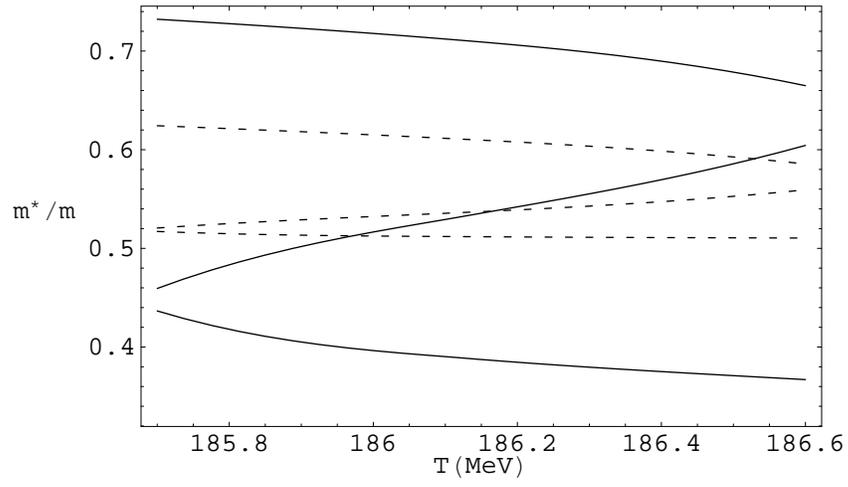
\item{{\bf Fig. 3}} The temperature  dependence of 
$ \frac{M^{*}_N}{M_N} $ and $\frac{ m^{*}_{\omega} }{ m_{\omega} }$ at 
$\mu=0$ in the range  $185.7$ MeV to $ 186.6 $ MeV where multiple
solutions for $M^*$ and $m_\omega^*$ exist.
The solid line represents $ \frac{M^{*}_N}{M_N} $ and the dashed line 
represents $ \frac{ m^{*}_{\omega} }{ m_{\omega} }  $.

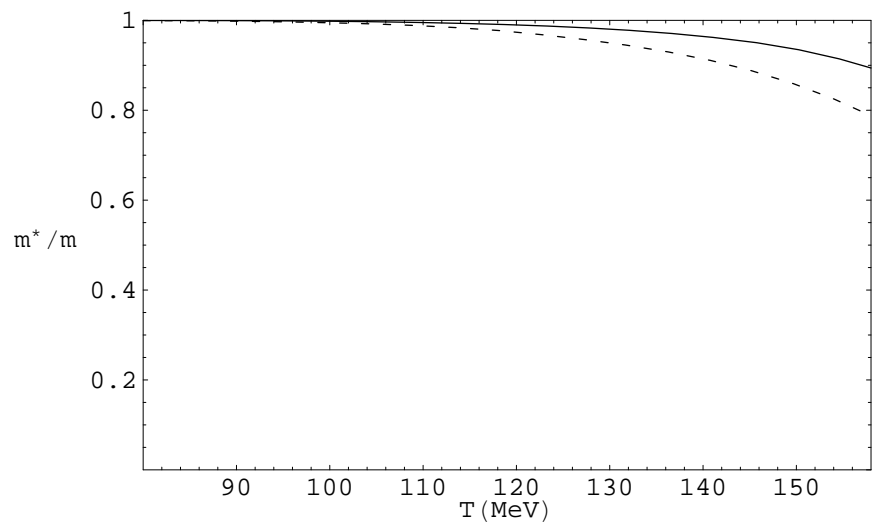
\item{{\bf Fig. 4}} The temperature  dependence of 
$ \frac{M^{*}_N}{M_N} $ and $\frac{ m^{*}_{\omega} }{ m_{\omega} }$
at $\mu=200$ MeV where  there is a single solution for $M^{*}$ and
$m^{*}_{\omega}$.    
The solid line represents $ \frac{M^{*}_N}{M_N} $ and the dashed line 
represents $ \frac{ m^{*}_{\omega} }{ m_{\omega} }  $.

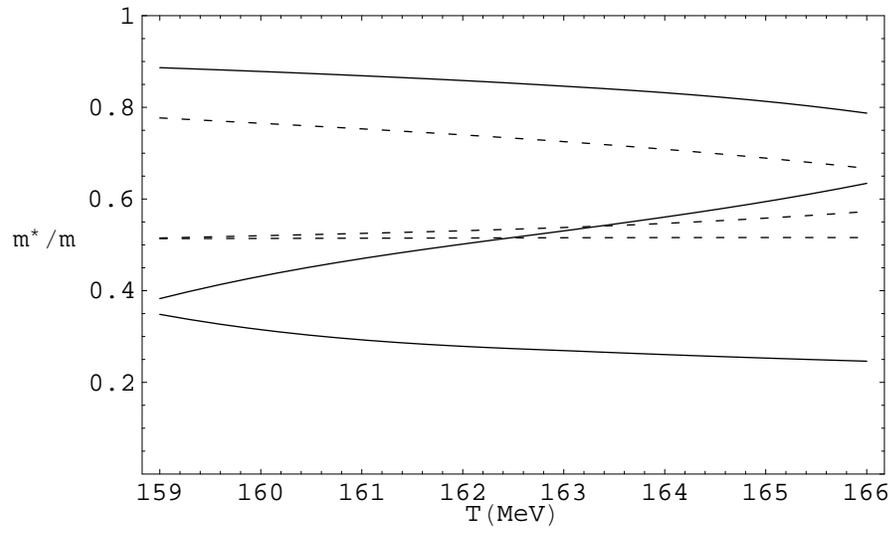
\item{{\bf Fig. 5}} The temperature  dependence of 
$ \frac{M^{*}_N}{M_N} $ and $\frac{ m^{*}_{\omega} }{ m_{\omega} }$ 
at $\mu=200$ MeV where there appear several solutions for $M^{*}$ and
$m^{*}_{\omega}$ and T ranges from $159$ MeV to $166$ MeV.The
  solid line represents $ \frac{M^{*}_N}{M_N} $ and the dashed line
represents $ \frac{ m^{*}_{\omega} }{ m_{\omega} }  $.

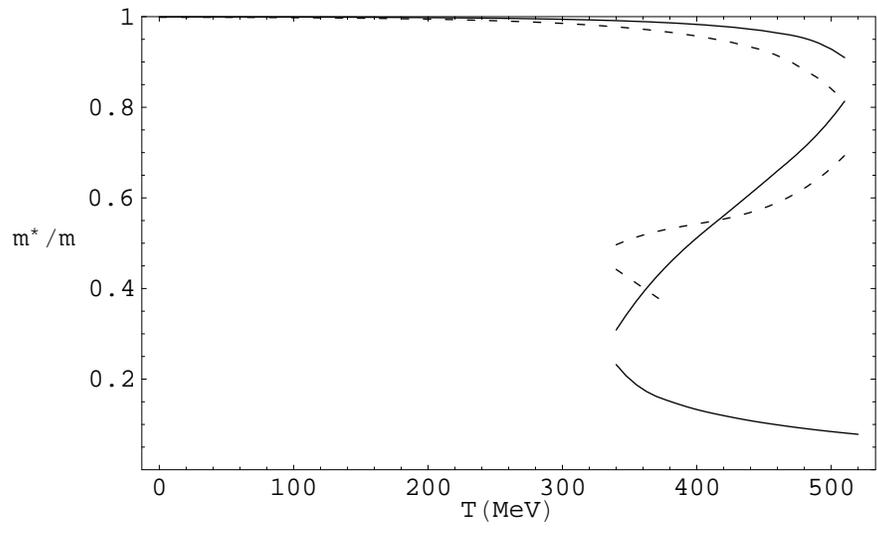
\item{{\bf Fig. 6}} The chemical potential dependence of 
$ \frac{M^{*}_N}{M_N}  $ and $ \frac{m^{*}_{\omega}}{ m_{\omega} }$ at
$ T=100$ MeV. The solid line represents $ \frac{M^{*}_N}{M_N} $
and the dashed line represents $ \frac{ m^{*}_{\omega} }{ m_{\omega} } $.

\item{{\bf Fig. 7}} The isotherms of the thermodynamic potential as a
function of $M^{*}$ at $\mu=0$. From top to bottom the five lines represent
the isothems of the thermodynamic potential at 
$T=180$ MeV, $185$ MeV, $186$ MeV,$ 186.3$ MeV and $ 188$ MeV respectively.

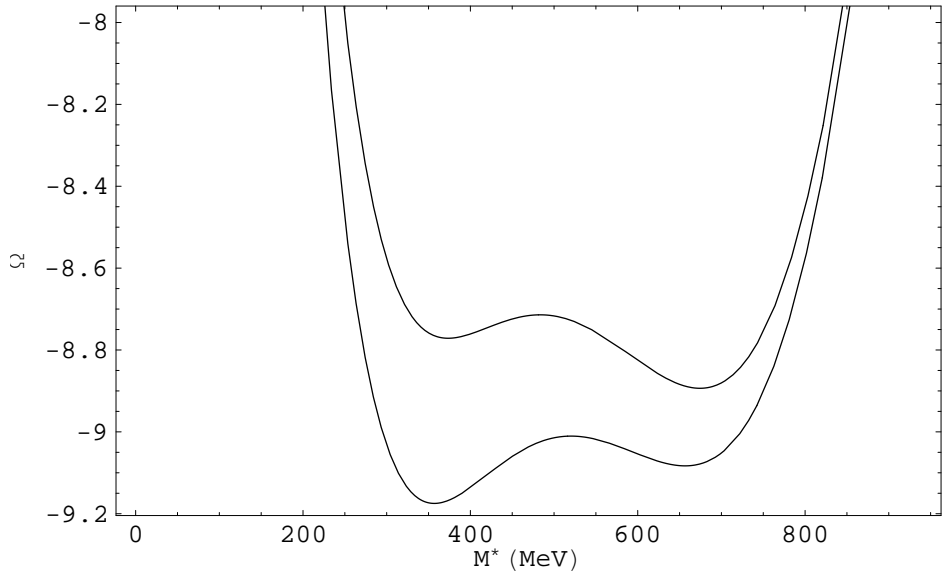
\item{{\bf Fig. 8}} An enlargement of the  isotherms of the 
thermodynamic potential $\Omega$ at $\mu=0$ when $\Omega$
has three extreme points at T $\sim 186$ MeV.

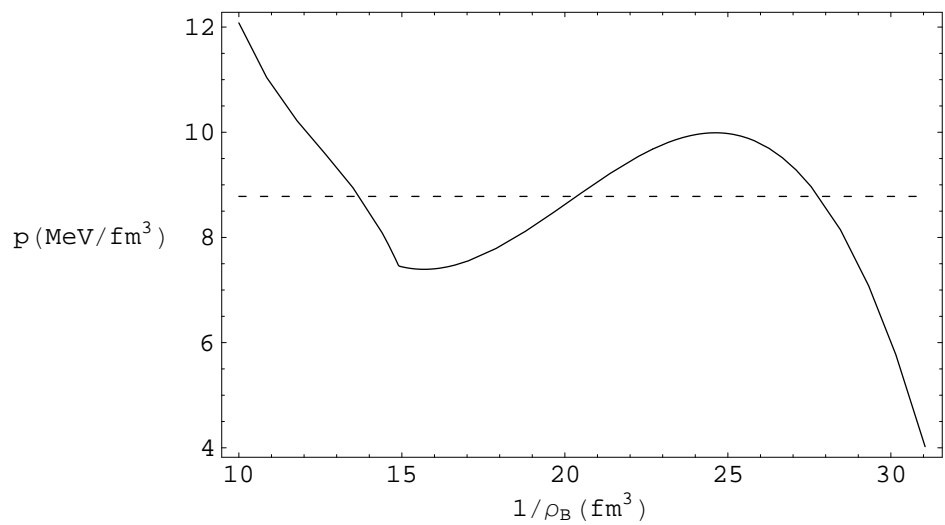
\item{{\bf Fig. 9}} The  phase diagram in nuclear matter: pressure as a function of
proper volume/baryon for $T=100$ MeV. The dashed line is Maxwell isotherm.
 
\end{description}

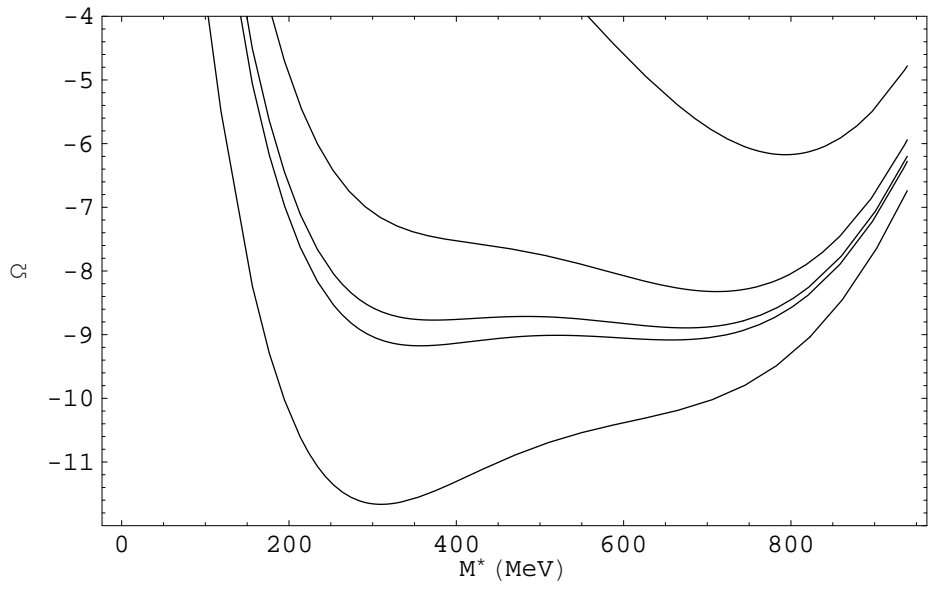
\end{document}